\documentclass[prl,twocolumn,showpacs,preprintnumbers,amsmath,amssymb]{revtex4}

\usepackage{graphicx}
\usepackage{dcolumn}
\usepackage{bm}

\begin{document}

\newcommand*{\cm}{cm$^{-1}$\,}


\title{Formation of the density wave energy gap in Na$_2$Ti$_2$Sb$_2$O: an optical spectroscopy study}

\author{Y. Huang}
\author{H. P. Wang}
\author{W. D. Wang}
\author{Y. G. Shi}
\author{N. L. Wang}

\affiliation{Beijing National Laboratory for Condensed Matter
Physics, Institute of Physics, Chinese Academy of Sciences,
Beijing 100190, China}


\begin{abstract}

We performed optical spectroscopy measurement on single crystals
of Na$_2$Ti$_2$Sb$_2$O, which is the parent compound for the newly discovered Ti-based superconductors.
The study reveals significant spectral change
across the phase transition at 114 K and formation of a density-wave type energy gap.
The opening of the gap removes most part of the free carrier spectral weight and causes 
a dramatic reduction of the carrier scattering rate. 
The ratio of 2$\Delta/k_BT_s\approx$14 is considerably larger than
the mean-field value based on the weak-coupling BCS theory.
The study also reveals a weak correlation effect in the titanium oxypnictides.
\end{abstract}



\maketitle

Charge- or spin-density waves (CDW/SDW) and superconductivity are
different collective quantum phenomena in solids. The interplay between density wave and superconductivity
has generated considerable interest in condensed matter physics. A recent example is the
iron-based superconductors where superconductivity emerges from the suppression
of the SDW instability \cite{Dong,Cruz}. It is widely believed that the superconductivity is mediated dominantly
by the spin fluctuations. The repulsive pairing interaction between the quasi-nested electron and hole Fermi
surfaces (FSs) leads to unconventional
superconductivity with a sign-changing (s$^{\pm}$-wave) pairing symmetry \cite{Mazin,Scalapino}.

Very recently, superconductivity in a titanium oxypnictide system Ba$_{1-x}$Na$_x$Ti$_2$Sb$_2$O has been
reported \cite{Yajima,Doan}. The undoped compound BaTi$_2$Sb$_2$O shows a superconducting transition
at T$_c$=1.2 K \cite{Yajima}. With Na doping, T$_c$ increases to 5.5 K \cite{Doan}.
Although the T$_c$ is very low, the system attracts much attention
because it shares a number of similarities to the Fe-based superconductors.
The compounds belong to a family with two-dimensional (2D) layered structures, consisting of
alternative stacking of conducting octahedral
layers Ti$_2$Pn$_2$O (Pn=As, Sb) and other insulating layers (e.g. Na$_2$,
Ba, (SrF)$_2$, (SmO)$_2$) (see inset of Fig. 1) \cite{Adam,Axtell,Ozawa1,Ozawa2,Ozawa3,Ozawa4,Liu1,Wang,Liu2,Zhai}.
Most notably, the compounds also show
competing phenomenon. The undoped compounds commonly show
phase transitions below certain temperatures. The earliest reported compound \cite{Adam} in this family,
Na$_2$Ti$_2$Sb$_2$O, shows a phase transition at \emph{T}$_s$=114 K as characterized
by a sharp jump in resistivity and a drop in spin susceptibility \cite{Ozawa1}.
Similar phase transitions at higher temperatures were also found
in the related compounds Na$_2$Ti$_2$As$_2$O (\emph{T}$_s$=320 K) \cite{Axtell,Ozawa1,Ozawa2,Ozawa3,Liu1}, BaTi$_2$As$_2$O (\emph{T}$_s$=200 K) \cite{Wang} and R$_2$Ti$_2$Pn$_2$O (R=(Sr,F), (Sm,O);
Pn=As, Sb; \emph{T}$_s$=198-380 K) \cite{Liu2}. In fact, the superconducting
BaTi$_2$Sb$_2$O also exhibits such a phase transition but at lower temperature \emph{T}$_s$=45 K \cite{Yajima}.
This transition temperature is further suppressed upon Na doping, by contrast the superconductivity is enhanced \cite{Doan}.

The microscopic mechanism for this phase transition has not been determined, but it has
been suggested to arise from the SDW or CDW instability driven by
the strong nested electron and hole Fermi surfaces \cite{Pickett,Biani,Singh,Yan,Subedi}. In particular,
a bicollinear or blocked checkerboard antiferromagnetic ordering
patterns have been predicted for the SDW ordered state for Na$_2$Ti$_2$Pn$_2$O (Pn=As, Sb)
from the first principle calculations \cite{Yan}. The presence
of nested electron and hole pockets and the antiferromagnetic instability would imply
a sign-changing s-wave symmetry in the superconducting titanium oxypnictides \cite{Singh},
similar to the Fe-based superconductors.

It is crucial to understand the origin of the phase transition in the undoped compound because it is
the essential step towards understanding the mechanism of superconductivity in doped systems. Up to now,
there is no spectroscopic experiment performed on those compounds. It is well known that, in the density
wave state (either CDW or SDW), the formation of electron-hole pairs with a nesting
wave vector connecting different regions of FSs would lead to
the opening of an energy gap \cite{DresselGruner}. This is a fundamental quantum phenomenon in solids, and
by that means, the system of interacting electrons can stabilize a broken symmetry ground state.
The precise measurement of the energy gap permits a meaningful
comparison to microscopic theories.

In this work we present optical spectroscopic study on Na$_2$Ti$_2$Sb$_2$O, which is a sister
compound to BaTi$_2$Sb$_2$O. Above the phase transition at 114 K, Na$_2$Ti$_2$Sb$_2$O is
quite metallic with rather high plasma frequency, $\omega_p\approx$ 20000 \cm (2.5 eV). Below the transition,
formation of density wave energy gap is clearly observed, leading to the removal of
most part of Drude component. Nevertheless, the FSs are only partially gapped because
residual itinerant carriers were still left. The residual Drude component becomes rather narrow and sharp,
indicating a dramatic reduction of the carrier
scattering rate. Our study reveals the ratio of 2$\Delta/k_BT_s \simeq$ 14, a value being considerably larger than
the mean-field value based on the weak-coupling BCS theory for a density wave phase transition.
Furthermore, the study also reveals weak correlation effect in the compound.

\begin{figure}
\includegraphics[width=3 in]{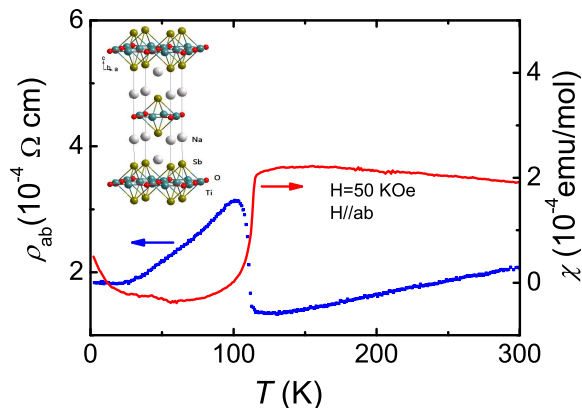}%
\vspace*{-0.20cm}%
\caption{\label{fig:R}(Color online) Temperature dependent in-plane resistivity 
and magnetic susceptibility of Na$_2$Ti$_2$Sb$_2$O. The inset
shows the crystal structure of Na$_2$Ti$_2$Sb$_2$O.}
\end{figure}

Plate-like single crystals of Na$_2$Ti$_2$Sb$_2$O with size up to 3$\times$3 mm$^2$
were grown from the NaSb flux method. Figure 1 shows the in-plane resistivity
and magnetic susceptibility measured in Quantum Design PPMS and SQUID-VSM, respectively.
The resistivity shows metallic behavior
at high temperature. A sharp upturn appears near 114 K. Then, metallic temperature
dependence is still observed at low temperature. At the transition, the susceptibility shows a sharp reduction.
A small upturn at low temperatures could be attributed to the presence of defects in the crystals.
A detailed description of growth and characterization of single crystals
in this family would be reported elsewhere \cite{Shi}.

The optical reflectance measurements were performed on a
combination of Bruker IFS 80v/s and 113v spectrometers in the frequency range
from 20 to 40000 cm$^{-1}$. An \textit{in situ} gold and aluminium
overcoating technique was used to get the reflectivity
R($\omega$). The real part of conductivity $\sigma_1(\omega)$ is
obtained by the Kramers-Kronig transformation of R($\omega$).

\begin{figure}
\includegraphics[width=2.7 in]{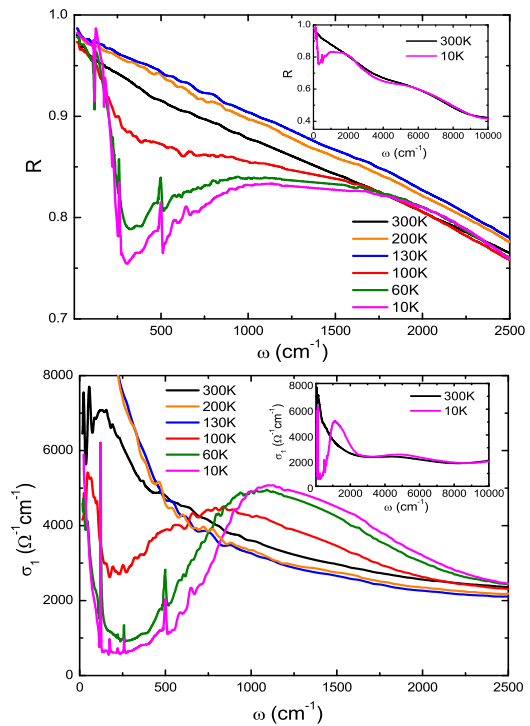}%
\vspace*{-0.20cm}%
\caption{\label{fig:R}(Color online) R($\omega$) and $\sigma_1(\omega)$ for Na$_2$Ti$_2$Sb$_2$O
below 2500 \cm. Insets: The reflectance and conductivity spectra over broad range of frequencies. }
\end{figure}

Figure 2 shows the low frequency
R($\omega$) (upper panel) and $\sigma_1(\omega)$ (lower panel) up to 2500 \cm. The insets show the spectra over broad frequency
region up to 10000 \cm. R($\omega$)
exhibits metallic response in both frequency and temperature dependences above the phase transition.
R($\omega$) displays roughly a linear-$\omega$ dependence and approaches to unity at zero
frequency. It also increases with decreasing temperature. The most prominent feature is a substantial suppression
in R($\omega$) for \emph{T}$<$\emph{T}$_s$, which is a strong
optical evidence for the formation of an energy gap. The
low-$\omega$ reflectance increases faster towards unity at zero
frequency than those at high \emph{T}. As a consequence, one can
see a rather sharp low-$\omega$ reflectance edge. This indicates
clearly that the Fermi surfaces are only partially gapped and the
compounds are still metallic below \emph{T}$_s$. The change of
R($\omega$) from an overdamped linear-$\omega$ dependent behavior
to a well-defined reflectance edge upon cooling the sample into
the ordered state immediately suggests a dramatic reduction of the
carrier scattering rate, while its low-energy location implies a
considerable reduction of carrier density. Besides the change in the electronic spectrum, five phonon
modes at 123, 175, 222, 260, 500 \cm emerge in R($\omega$) at low T. This is due
to the substantially reduced screening effect caused by the reduction of itinerant carriers.

The evolution of the electronic states is more clearly
reflected in the conductivity spectra. At high temperatures, broad Drude-like features
are observed in $\sigma_1(\omega$). Below \emph{T}$_s$,
the spectra at low frequencies are severely
suppressed. The suppressed spectral weight is transferred mostly to the region above the gap,
resulting in a pronounced peak structure near 1100 \cm. Associated
with the low-$\omega$ reflectance edge, a very sharp and narrow
Drude component emerges below this peak. The $\sigma_1(\omega)$ data show more clearly the partial
gapping of the FSs. It is known that the hallmark of a symmetry-broken phase transition, such as
superconductivity or density wave order, is the formation of an energy gap
near the Fermi level E$_F$, resulting
in the lowering of the total energy of the system. However, due to different coherent
factors (case I for density wave and case II for superconductivity) \cite{DresselGruner}, the characteristic energy
gap features of superconductivity and density wave orders are different in optical conductivity. In
an s-wave superconducting state at $T=0$, the absorption smoothly
rises at the gap frequency and gradually merges to the conductivity spectrum at high frequency,
while for a density wave order, the opening of an energy gap
leads to a pronounced peak
just above the energy gap in $\sigma_1(\omega)$ \cite{DresselGruner,Degiorgi,Hu}.
The observation of a characteristic peak structure above the gap yields optical
evidence for a density wave type phase transition.

A notable point is that the gap size at the lowest measurement 
temperature (10 K), which could be identified as the peak position at 1100 \cm, is
surprisingly large. The ratio of the energy gap relative to the transition temperature
2$\Delta/k_BT_s$ is about 14, a value being much larger than
the BCS mean field value of 3.5 for a density wave phase transition \cite{DresselGruner}. This 
means that the transition temperature is significantly lower than the mean field transition 
temperature. The reason is unclear. It might be due to highly two-dimensional electronic structure 
which leads to strong fluctuation effect and suppresses the actual ordering temperature. Similar gap ratios 
were also observed in other low-dimensional density wave materials, 
e.g. (TaSe$_4$)$_2$I \cite{DresselGruner}. Nevertheless, further studies is necessary to resolve this issue.

\begin{figure}
\includegraphics[width=3 in]{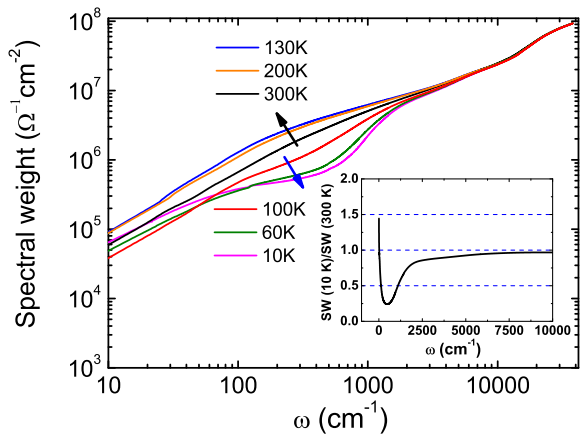}%
\vspace*{-0.20cm}%
\caption{\label{fig:R}(Color online) Frequency-dependent spectral weight for Na$_2$Ti$_2$Sb$_2$O
at different temperatures. The two arrows indicate different trends for the temperatures
above and below the phase transitions. Inset: the normalized spectral weight
SW(10 K)/SW(300 K) up to 10000 \cm.}
\end{figure}

The above mentioned spectral weight transfer is also seen clearly
in the spectral weight plot shown in Fig. 3. The
spectral weight is defined as
SW=$\int^{\omega_c}_0\sigma_1(\omega)d\omega$, where
$\omega_c$ is a cut-off frequency. Above the phase transition,
the low-$\omega$ spectral weight increases with decreasing temperature due to the
narrowing of the Drude component, however, below $T_s$, the spectral
weight is severely suppressed below 1100 \cm owing to the formation of energy gap. The inset
shows the plot of the normalized spectra SW(10 K)/SW(300 K). At low-frequency, SW(10
K)/SW(300 K) is somewhat higher than unity, then drops sharply with increasing frequency. This is
due to the presence of sharp residual Drude component at 10 K. The normalized
spectral weight reaches a minimum near 500 \cm, then
recovers quickly at higher frequencies. It can be ascribed to the density wave
energy gap formation, resulting in a transfer of the spectral weight to the region just above the energy gap. 
The small spectral weight recovery at higher energy scale above 3000 \cm could be attributed to the
weak temperature dependent interband transitions. Those interband transitions
would involve the bands across the Fermi level. Due to the effect of temperature-dependent
Fermi distribution function, the electron occupations near the Fermi level on those bands
could show a small change at different temperatures.

To quantitatively characterize the spectral change, particularly the evolution of the Drude
part, across the phase transition, we decompose the optical conductivity
spectral into different components using a Drude-Lorentz analysis.
The dielectric function has the form \cite{Hu}
\begin{equation}
\epsilon(\omega)=\epsilon_\infty-\sum_{i}{{\omega_{p,i}^2}\over{\omega_i^2+i\omega/\tau_i}}+\sum_{j}{{\Omega_j^2}\over{\omega_j^2-\omega^2-i\omega/\tau_j}}.
\label{chik}
\end{equation}
where $\epsilon_\infty$ is the dielectric constant at
high energy, and the middle and last terms are the Drude and Lorentz components,
respectively. The Drude components represent the contribution from itinerant
electrons, while the Lorentz components describe the excitations across the gap and interband
transitions. It is found that, in order to reproduce the low-$\omega$ conductivity
reasonably well, two Drude components, a narrow one and a broad one,
have to be used. Since the compound has multiple FSs, the inclusion of two Drude-components 
appears to be natural. The broad Drude component takes most part of the spectral weight of the
itinerant electrons, while the narrow one only occupies a small fraction of this spectral weight.

\begin{figure}
\includegraphics[clip,width=2.7in]{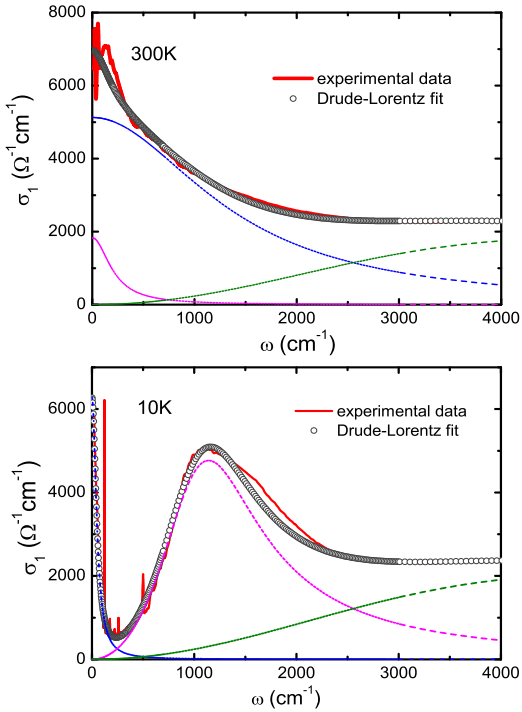}
\caption{(Color online) The experimental data of $\sigma_1(\omega)$ at 300 K
and 10 K together with the Drude-Lorentz fits shown at the bottom.}
\end{figure}

Figure 4 shows the $\sigma_1(\omega)$ at 300
K and 10 K together with the Drude-Lorentz fitting components. 
Interestingly, we find that, in the ordered state below \emph{T}$_s$,
the gapping of the FSs mainly removes the spectral weight of the broad Drude
component. There is a little change for the spectral weight of narrow Drude component,
but its width is significantly reduced. The results suggest that the gapping of 
the FSs caused by the density wave phase transition mainly occurs on 
those FSs where the electrons experience stronger scattering. In the mean time,
due to the removal of those FSs, the electrons
in the residual FSs become further less scattered.

Since two Drude components contribute to the conductivity, the
overall plasma frequency $\omega_p$ could be calculated as
$\omega_p$=($\omega_{p1}^2$+$\omega_{p2}^2$)$^{1/2}$. Then, we get $\omega_p\approx$20000 \cm (2.5 eV)
for all measurement temperatures above \emph{T}$_s$. 
This value is dramatically reduced in the ordered state. At 10 K, only a very narrow Drude component is
left, leading to $\omega_p\approx$ 4300 \cm (0.53 eV).
The overall plasma frequency could also be estimated by summarizing the low-$\omega$ spectral weight,
$\omega_p^2$=8$\int^{\omega_c}_0\sigma_1(\omega)d\omega$. The integration
up to $\omega_c$ should cover all the spectrum contributed by the free
carriers but still below the inter-band transition. For $T>T_s$, we take $\omega_c$=3000 \cm where we
expect that there is a balance between the Drude component tail
and the onset part of the interband transition. 
Then, we also get $\omega_p\approx$20000 \cm. The spectral
weight of the Drude component in the low-T ordered
state could be more accurately determined because it separates
distinctly from the remaining part in the conductivity spectra. For example,
we take $\omega_c$=220 \cm at T=10 K, which leads to $\omega_p\approx$ 4300 \cm. Therefore, we get
essentially the same values of overall plasma frequencies as from Drude-Lorentz analysis.
Since, effectively, the square of the plasma frequency $\omega_p^2$ is proportional
to \emph{n/m$_{eff}$} (where \emph{n} is the
carrier density, \emph{m$_{eff}$} is the effective mass), if assuming that the
effective mass of itinerant carriers would not change with
temperature, then the residual carrier density is only 4.6$\%$ of
that at high temperature. This means that
roughly over 95$\%$ of FSs are removed associated with the density wave phase
transition.

Determination of the plasma frequency $\omega_p$ helps to elucidate
the correlation effect by comparison with the result from
band theory calculations. Within a single-band tight binding
model, the kinetic energy energy of the electrons is proportional to
the square of the plasma frequency. Then, the ratio of the
experimental kinetic energy and the theoretical kinetic energy
from band structure calculations could be calculated from the plasma frequencies
determined respectively by experiments and band structure calculations,
\textit{K}$_{exp}$/\textit{K}$_{band}$=$\omega_{p,exp}^2/\omega_{p,band}^2$. It
provides a measure for the band renormalizations from the
correlation effect \cite{Qazilbash,Qimiao}. \textit{K}$_{exp}$/\textit{K}$_{band}$ is
close to unity for a simple metal, but is reduced to zero for a
strongly correlated Mott insulator. The band
structural calculations give the ab-plane plasma frequency of
3.0 eV in the paramagnetic phase \cite{Pickett}. Then we can estimate that,
\textit{K}$_{exp}$/\textit{K}$_{band}$=$\omega_{p,exp}^2/\omega_{p,band}^2$$\approx$ 0.7.
In Fe-pnictides, the value of this ratio is close to 0.4 \cite{Qazilbash,Qimiao,ZGChen}.
Therefore, the obtained value suggests that
the correlation effect in Na$_2$Ti$_2$Sb$_2$O is substantially
weaker than iron-pnictides.

The present work firmly establishes the density wave origin for the phase transition in the
titanium oxypnictide Na$_2$Ti$_2$Sb$_2$O. However, we can not distinguish whether the ordered state is
a CDW or an SDW order for the reason that both broken symmetry states have
the same coherent factor. Since both orders have been predicted by the
density function calculations on BaTi$_2$Sb$_2$O and related compounds \cite{Pickett,Biani,Singh,Yan,Subedi},
it is crucial to determine the order experimentally. The earlier neutron diffraction measurement on
Na$_2$Ti$_2$Sb$_2$O did not reveal a magnetic order at low T \cite{Ozawa1}. However, there is a
possibility that the ordered moment is
very small and below the resolution limit of the measurement. The magnetic susceptibility data displayed in
Fig. 1 show relatively large values at high T and sharp drop at transition which seems to favor an SDW origin.
Nevertheless, more sensitive experimental probes, for
example, $\mu$SR, should be used to detect any possible magnetic order in the compounds.
The issue is related to the pairing mechanism for the superconducting samples in the family. The SDW origin of
the instability would favor an unconventional superconductivity with sign-changing s-wave pairing, while
the CDW origin would suggest more conventional superconductivity with simple s-wave pairing.

To summarize, the ab-plane optical measurements of Na$_2$Ti$_2$Sb$_2$O
single crystals were performed. The study revealed dramatic spectral change
across the phase transition at 114 K and formation of a density-wave energy gap at low T.
The opening of the gap removes most part of the free carrier spectral weight and causes
a substantial reduction of the carrier scattering rate. 
The ratio of 2$\Delta/k_BT_s\approx$14 well exceeds the weak-coupling BCS value of 3.52, suggesting
that the transition temperature is significantly lower than the mean-field transition temperature. 
The study also revealed a weak correlation effect in the titanium oxypnictides.

\begin{acknowledgments}
We acknowledge helpful discussions with X. H. Chen, Z. Y. Lu, J. L. Luo and Q. M. Si.
This work was supported by the National Science Foundation of
China, and the 973 project of the Ministry of Science and Technology of China (2011CB921701,2012CB821403).
\end{acknowledgments}


\begin{thebibliography}{20}

\bibitem{Dong} J. Dong, H. J. Zhang, G. Xu, Z. Li, G. Li, W. Z. Hu, D. Wu,
G. F. Chen, X. Dai, J. L. Luo, Z. Fang, and N. L. Wang, Europhys.
Lett. \textbf{83}, 27006 (2008).


\bibitem{Cruz} Clarina de la Cruz, Q. Huang, J. W. Lynn, Jiying Li,
W. Ratcliff II, J. L. Zarestky, H. A. Mook, G. F. Chen, J. L. Luo,
N. L. Wang, and Pengcheng Dai, Nature \textbf{453}, 899 (2008).

\bibitem{Mazin} I. I. Mazin, D. J. Singh, M. D. Johannes, and M. H. Du,
Phys. Rev. Lett. \textbf{101}, 057003 (2008).

\bibitem{Scalapino} D. J. Scalapino, Rev. Mod. Phys. \textbf{84}, 1383 (2012).


\bibitem{Yajima} T. Yajima, K. Nakano, F. Takeiri, T. Ono, Y. Hosokoshi,
Y. Matsushita, J. Heister, Y. Kobayashi, and H. Kageyama, J. Phys. Soc. Jpn. \textbf{81}, 103706 (2012).

\bibitem{Doan} P. Doan, M. Gooch, Z. Tang, B. Lorenz, A. Moeller, J.
Tapp, P. C. W. Chu, and A. M. Guloy, J. Am. Chem. Soc. \textbf{134}, 16520 (2012).

\bibitem{Adam} A. Adam and H.-U. Schuster, Z. Anorg. Allg. Chem. \textbf{584}, 150
(1990).

\bibitem{Axtell} E. A. Axtell, III, T. Ozawa, S. M. Kauzlarich, and R. R.
P. Singh, J. Solid State Chem. \textbf{134}, 423 (1997).

\bibitem{Ozawa1} T. C. Ozawa, T. Pantoja, E. A. Axtell, III, S. M. Kau-
zlarich, J. E. Greedan, M. Bieringer, and J. W. Richardson
Jr., J. Solid State Chem. \textbf{153}, 275 (2000).

\bibitem{Ozawa2} T. C. Ozawa, S. M. Kauzlarich, M. Bieringer, and J. E.
Greedan, Chem. Mater. \textbf{13}, 1804 (2001).

\bibitem{Ozawa3} T. C. Ozawa and S. M. Kauzlarich, J. Cryst. Growth
\textbf{265}, 571 (2004).

\bibitem{Ozawa4} T. C. Ozawa and S. M. Kauzlarich, Sci. Technol. Adv.
Mater. \textbf{9}, 033003 (2008).

\bibitem{Liu1} R. H. Liu, D. Tan, Y. A. Song, Q. J. Li, Y. J. Yan, J. J.
Ying, Y. L. Xie, X. F. Wang, and X. H. Chen, Phys. Rev.
B \textbf{80}, 144516 (2009).

\bibitem{Wang} X. F. Wang, Y. J. Yan, J. J. Ying, J. Q. Li, M. Zhang,
N. Xu, X. H. Chen, J. Phys.: Condens. Matter \textbf{22}, 075702
(2010).

\bibitem{Liu2} R. H. Liu, Y. A. Song, Q. J. Li, J. J. Ying, Y. J. Yan, Y.
He, and X. H. Chen, Chem. Mater. \textbf{22}, 1503 (2010).

\bibitem{Zhai} H. F. Zhai, W. H. Jiao, Y. L. Sun, J. K. Bao, H. Jiang, X. J. Yang, 
Z. T. Tang, Q. Tao, X. F. Xu, C. Cao, J. H. Dai, Z. A. Xu, and G. H. Cao, arXiv:1211.4548.

\bibitem{Pickett} W. E. Pickett, Phys. Rev. B \textbf{58}, 4335 (1998).

\bibitem{Biani} F. F. de Biani, P. Alemany, and E. Canadell: Inorg. Chem.
\textbf{37}, 5807 (1998).

\bibitem{Singh} D. J. Singh, New J. Phys. \textbf{14}, 123003 (2012).

\bibitem{Yan} X. W. Yan and Z. Y. Lu, arXiv 1210.3481.

\bibitem{Subedi} A. Subedi, arXiv 1210.0499.


\bibitem{DresselGruner} M. Dressel and G. Gr\"{u}ner, "Electrodynamics of
Solids: Optical Properties of Electrons in Matter", Cambridge
University Press, (2002).

\bibitem{Shi} Y. G. Shi, H. P. Wang, W. D. Wang, Y. Huang, and N. L. Wang, unpublished.

\bibitem{Degiorgi} L. Degiorgi et al., Phys. Rev. Lett. \textbf{76}, 3838 (1996).

\bibitem{Hu} W. Z. Hu, J. Dong, G. Li, Z. Li, P. Zheng, G. F. Chen, J. L. Luo, and N. L.
Wang, Phys. Rev. Lett. \textbf{101}, 257005 (2008).

\bibitem{Qazilbash} M. M. Qazilbash, J. J. Hamlin, R. E. Baumbach, Lijun
Zhang, D. J. Singh, M. B. Maple, D. N. Basov, Nature Physics
\textbf{5}, 647 (2009).

\bibitem{Qimiao} Qimiao Si, Nature physics \textbf{5}, 629 (2009).


\bibitem{ZGChen} Z. G. Chen, R. H. Yuan, T. Dong, N. L. Wang, Phys. Rev. B \textbf{81}, 100502(R)
(2010).




\end{thebibliography}

\end{document}